\documentclass[prc,twocolumn,superscriptaddress,showpacs,showkeys,twoside,
               floatfix]{revtex4}

\usepackage{amssymb,epsfig}

\begin{document}

\title{Three-body resonant radiative capture reactions in astrophysics}

\author{L.\ V.\ Grigorenko}
\affiliation{Pervomayskaya st., 43, 52, RU-105043, Moscow, Russia}
\affiliation{Russian Research Center ``The Kurchatov Institute'', RU-123182
Moscow, Russia}

\author{M.\ V.\ Zhukov}
\affiliation{Department of Physics, Chalmers University of Technology
and G\"{o}teborg University, S-41296 G\"{o}teborg, Sweden}

\date{\today.}

\begin{abstract}
We develop the formalism based on the S-matrix for $3 \rightarrow 3$ scattering
to derive the direct three-body resonant radiative capture reaction rate.
Within this formalism the states, which decay only/predominantly directly into
three-body continuum, should also be included in the capture rate calculations.
Basing on the derivation, as well as on the modern experimental
data and theoretical calculations concerning $^{17}$Ne nucleus, we
significantly update the reaction rate for $^{15}$O($2p$,$\gamma$)$^{17}$Ne
process in explosive environment. We also discuss possible implementations for
the $^{18}$Ne($2p$,$\gamma$)$^{20}$Mg, $^{38}$Ca($2p$,$\gamma$)$^{40}$Ti, and
$^4$He($n\alpha$,$\gamma$)$^{9}$Be reactions.
\end{abstract}

\pacs{ 21.45.+v -- Few-body systems, 26.30.+k -- Nucleosynthesis in novae,
supernovae and other explosive environments, 25.40.Lw -- Radiative capture,
25.40.Ny -- Resonance reactions.}

\keywords{
three-body resonant radiative capture reactions,
$rp$ process waiting points,
hyperspherical harmonic method}

\maketitle


\section{Introduction}


The reactions of the three-body radiative capture may play a considerable role
in the rapid nuclear processes which take place in stellar media under the
conditions of high temperature and density.
The possibility to
bridge the waiting points of the $rp$-process in the explosive hydrogen burning
by the $2p$ radiative capture reactions was discussed in Ref.\ \cite{gor95}.
The reactions
$^{15}$O($2p$,$\gamma$)$^{17}$Ne, $^{18}$Ne($2p$,$\gamma$)$^{20}$Mg, and
$^{38}$Ca($2p$,$\gamma$)$^{40}$Ti could be a more
efficient way to ``utilize'' $^{15}$O, $^{18}$Ne, and $^{38}$Ca, than to wait
for their $\beta^+$ decay (corresponding lifetimes are 122, 1.67, and 0.44
seconds).
The $^4$He($n\alpha$,$\gamma$)$^{9}$Be
reaction has been found to be important for building heavy elements in the
explosions of supernovae \cite{woo94,tak94}. This reaction has been
several times theoretically considered in the recent years
\cite{fow75,gor95a,cau88,efr98,ang99,buc01}.

The three-body radiative capture is a very improbable process. It can only be
important if the sequence of the two-body radiative captures to the bound states
is not possible. This happens if the bound intermediate system does not exist
(along the driplines the continuum ground states are not uncommon). The
three-body radiative capture can proceed {\em sequentially} via the intermediate
resonances or {\em directly} from the three-body continuum. The later is the
inverse process to the ``true'' two-proton radioactivity \cite{gol60}, which
studies are active now
\cite{gri00,gri01,gri02,pfu02,gio02,chr02,gri03,gri03a,bro03,gri03b}. The
relations
between sequential and direct mechanisms of two-proton decay are discussed in
details in Refs.\ \cite{gri01,gri03a}.

In the modern literature exists some misunderstanding about the role of the
direct three-body capture in the theoretical calculations of the three-body
radiative capture rates. In the most cases this misunderstanding does not lead
to any significant problems. However, in some situations the difference
is sufficiently large.
In our opinion the origin of the misunderstanding is the following. The accurate
formulae for the resonant three-body capture are known for a long time (see
e.g.\ Ref.\ \cite{fow67}). But nowadays they does not seem to be always
interpreted completely correctly. The possible reason is that for nonresonant
capture calculations the sequential capture formalism is used
\cite{nom85,gor95,gor95a,ang99,buc01}. At some stage it has become
considered as obtained in more general assumptions (see e.g.\ Ref.\
\cite{nom85}) than the derivation of Ref.\ \cite{fow67}, based on complete
thermal equilibrium and detailed balance.

In this paper we use formalism based on the S-matrix for $3 \rightarrow 3$
scattering to derive the reaction rates for the three-body resonant radiative
capture. In this approach the right way of using these formulae becomes evident.
We find that the direct and sequential capture mechanisms are complementing each
other. In the cases when the sequential process is prohibited energetically or
suppressed dynamically the sequential formalism should underestimate the rate.
Among the processes, where significant differences with previous calculations
can be found are reactions leading to $^{17}$Ne and $^{40}$Ti.

The unit system $\hbar=c=1$ is used in the article.


\section{Three-body radiative capture}


The derivations provided below are relatively trivial. Those of Section
\ref{sec:seq-cap} can be found e.g.\ in Refs.\ \cite{nom85,gor95}. They are
presented, however, in much details to provide unified notations, simplify
the reading of the paper and to avoid any possible misinterpretation.


\subsection{Sequential capture}

\label{sec:seq-cap}

The abundance $Y_{A+2}$ for the nucleus with mass number $A+2$ due to the
sequential two-proton capture reaction on the nucleus with mass number $A$ is
defined via three-body reaction rate
$\left\langle \sigma_{pp,\gamma}v\right\rangle$ as (see e.g.\ Ref.\ 
\cite{fow67}):
\begin{equation}
\dot{Y}_{A+2}=(1/2) \; N_{A}^{2}\,\rho^{2}\;
\left\langle \sigma_{pp,\gamma}v\right\rangle \;Y_{p}^{2}Y_{A},
\end{equation}
where $\rho$ is density and $N_A$ is Avogadro number.
The two-proton reaction rate is defined for sequential capture of protons (for
example in \cite{gor95}) by
\[
\left\langle \sigma_{pp,\gamma}v\right\rangle =
2 \sum_{i}\frac{\left\langle \sigma_{p,p}v\right\rangle _{i}}{\Gamma_{p,i}}
\left\langle \sigma_{p,\gamma
}v\right\rangle _{i}\;.
\]
This expression is a consequence of the rate equations for the resonance
number $i$:
\begin{eqnarray}
\dot{Y}_{A+1}^{(i)}  & =&N_{A}\,\rho\;\left\langle \sigma_{p,p}v\right\rangle
_{i}\;Y_{p}Y_{A}-\Gamma_{p,i}Y_{A+1}^{(i)} \nonumber  \\
\dot{Y}_{A+2}  & =&\sum_{i}N_{A}\,\rho\;\left\langle \sigma_{p,\gamma
}v\right\rangle _{i}\;Y_{p}Y_{A+1}^{(i)}
\label{eq:sys-eq}
\end{eqnarray}
and the assumption about thermodynamic equilibrium for the intermediate
resonant states $\dot{Y}_{A+1}^{(i)}=0$.

The standard expression for cross section of the resonance reaction
with entrance channel $\alpha$ and exit channel $\beta$ is
\begin{equation}
\sigma(E)=\frac{\pi}{k_{12}^{2}}\,
\frac{\Gamma_{\alpha}\Gamma_{\beta}}{(E-E_{R})^{2}+\Gamma^{2}/4} \,
\frac{2J_{2R}+1}{(2J_{1}+1)(2J_{2}+1)} \, ,
\end{equation}
where $J_{1}$ and $J_{2}$ are the total spins of incoming particles and
$J_{2R}$ is the total spin of the resonance.

In the case of intermediate capture into the narrow proton resonance
number $i$ (which also decays practically only via proton emission)
$\Gamma_{\alpha}=\Gamma_{\beta}=\Gamma_{p,i}\,$, and
\begin{eqnarray}
\left\langle \sigma_{p,p}v\right\rangle _{i}  =
\int v \, \sigma_i(E_{12}) \, w(k_{12}) \, d^3k_{12} =
\left(   \frac{A_{1}+A_{2}}{A_{1}A_{2}}\right)^{3/2}
 \nonumber \\
\times  \; \frac{2J_{2R,i}+1}{2(2J_{I}+1)} \left( \frac{2\pi}{m k T}\right)
^{3/2} \! \exp\left[-\frac{E_{2R,i}}{k T} \right]  \Gamma_{p,i} \, , \quad
\end{eqnarray}
where $J_{I}$ is the total spin of the initial (core) nucleus and $J_{2R,i}$
is the total spin of the resonance number $i$ in the core+$p$ system. The
Boltzmann distribution over the relative motion momentum
$k_{12}=\sqrt{2m_{12}E_{12}}$ is
\[
w(k_{12})={(2\pi m_{12}kT)^{-3/2} \exp[  -E_{12}/kT]} \, ,
\]
and we approximate the integral over the resonance profile as
\[
\int_{-\infty}^{\infty}\frac{dE}{(E-E_{R})^{2}+\Gamma^{2}/4} =
\frac{2\pi}{\Gamma} \, .
\]
The reaction rate for the subsequent capture of the second proton on the system
core+$p$ in the resonant state number $i$ and the following gamma emission is:
\begin{eqnarray}
\left\langle \sigma_{p,\gamma}v\right\rangle _{i} & = &
\left(  \frac{A_{1}+A_{2}+A_{3}}{(A_{1}+A_{2})A_{3}}\right)^{3/2}\!\!\!\!
\frac{2J_{F}+1}{2(2J_{2R,i}+1)} \left(\frac{2\pi}{m k T}\right)^{3/2} \nonumber
\\
&\times & \, \exp\left[  -(E_{3R}-E_{2R,i})/kT \right] \;
\Gamma_{\gamma} \Gamma_{p,i}^{\prime} / \Gamma_{3R} \, ,
\end{eqnarray}
where $J_{F}$ is the total spin of the resonance $E_{3R}$ in the core+$2p$
system. $\Gamma_{p,i}^{\prime}$ is the partial width for decay of this state
into the binary channel (core+$p$)+$p$, where the core+$p$ system is in the
resonant state number $i$. It is easy to find out that the two-proton
reaction rate for sequential proton capture (through narrow intermediate
resonances) is
\begin{eqnarray}
\left\langle \sigma_{pp,\gamma}v\right\rangle & = &
\left( \frac{A_{1}+A_{2} + A_{3}}{A_{1}A_{2}A_{3}}\right)^{3/2} \!\!
\frac{2J_{F}+1}{2(2J_{I}+1)} \left(\frac{2\pi}{m k T}\right)^{3} \nonumber \\
&\times & \, \exp\left[ -E_{3R}/kT \right] \;
\textstyle \left(
\Gamma_{\gamma}\sum_{i} \Gamma_{p,i}^{\prime}\right)/\Gamma_{3R}\, .
\label{eq:prod-rate-seq-1}
\end{eqnarray}
The following features of the rate Eq.\ (\ref{eq:prod-rate-seq-1}) should be
noted:

\noindent (i) The reaction rate does not depend on the number and properties of
the intermediate states, but only on the sum of the proton widths for population
of these states.

\noindent (ii) In the most expected case of the sequential decay
mode dominance for the three-body resonance $E_{3R}$ we have
$\Gamma_{\gamma}\ll\Gamma_{3R}$, $\Gamma_{2p}\ll\Gamma_{3R}$ ($\Gamma_{2p}$ is
the width for the {\em direct} decay into $2p$ continuum, the process not
proceeding via intermediate core+$p$ resonances) and
$\Gamma_{3R}=\sum_{i}\Gamma_{p,i}^{\prime}\,$.
So, the reaction rate depends {\em only} on gamma width of the three-body
resonance.

\noindent (iii) Eq.\ (\ref{eq:prod-rate-seq-1}) shows that
if there exist other significant decay channels for three-body resonance
$E_{3R}$, then $\Gamma_{3R}>\sum_{i}\Gamma_{p,i}^{\prime}$
and the production rate decreases. Such
possible decay channel is, already mentioned, direct (not via resonances)
decay of the three-body resonance $E_{3R}$
into two-proton continuum. Typically this process is suppressed, but there are
cases where this process is not suppressed. It can be dominating, or even it
can be  the only
possible decay channel (no intermediate resonances). Such an opportunity is
considered in the next Section.


\subsection{Direct capture}

\label{sec:dir-capt}

The abundance  $Y_{A+2}$ due to the direct two-proton capture reaction is
defined via three-body reaction rate as
\begin{equation}
\dot{Y}_{A+2}=(1/2) \; N_{A}^{2} \,\rho^{2}\;\left\langle
\sigma_{2p,\gamma}v\right\rangle
\;Y_{p}^{2}Y_{A}.
\label{eq:abund}
\end{equation}
To derive the cross section of the direct capture from the three-body continuum
we use the S-matrix formalism for $3\rightarrow3$ reaction. The plane wave for
three particles can be decomposed over hyperspherical harmonics
$\mathcal{I}_{K\gamma}^{LM_L}$:
\begin{eqnarray}
\Psi_{3}^{pw}=\exp[\,i\mathbf{k}_{1}\mathbf{r}_{1}+i\mathbf{k}_{2}\mathbf{r}_{2}
+i\mathbf{k}_{3}\mathbf{r}_{3}] \, \chi_{S_1 M_1}\chi_{S_2 M_2}\chi_{S_3 M_3}
\nonumber \\
=\exp[\,i\mathbf{k}_{cm}\mathbf{R}_{cm}
+i\mathbf{k}_{y}\mathbf{Y}+i\mathbf{k}_{x}\mathbf{X}]
\, \chi_{S_1 M_1}\chi_{S_2 M_2}\chi_{S_3 M_3}\nonumber \\
=\; \exp[\,i\mathbf{k}_{cm}\mathbf{R}_{cm}] \;
\frac{\left(  2\pi\right)  ^{3}} {\left(  \varkappa \rho\right) ^{2}}
\sum_{JM} \, \sum_{KLl_xl_ySS_x}  i^{K}J_{K+2}(\varkappa\rho) \nonumber \\
\times \; \mathcal{J}_{KLl_xl_ySS_x}^{JM}(\Omega_{\rho}) \, \sum_{M_L}
\mathcal{I}_{Kl_xl_y}^{LM_L\ast}(\Omega_{\varkappa}) \;
g^{JLSS_x}_{MM_LM_1M_2M_3}\, ,
\label{eq:pw3}
\end{eqnarray}
where $\chi_{S_iM_i}$ are spin functions for the cluster number $i$.
$\mathcal{J}^{JM}$ denote the hyperspherical harmonic $\mathcal{I}^{LM_L}$
coupled with spin functions of clusters to the total momentum $J$
\begin{eqnarray}
\mathcal{J}_{KLl_xl_ySS_x}^{JM} =
[\mathcal{I}_{Kl_xl_y}^{L} \otimes X_{SS_x}]_{JM}\, ,
\nonumber  \\
X_{SS_xM_S}=[[\chi_{S_1} \otimes \chi_{S_2}]_{S_x} \otimes \chi_{S_1}]_{SM_S}
\, . \nonumber
\end{eqnarray}
Variables $\mathbf{k}_{cm}$ and $\mathbf{R}_{cm}$ describe the center of mass
motion; $\{\mathbf{k}_x,\mathbf{k}_y\}$ and $\{\mathbf{X},\mathbf{Y}\}$ are
conjugated sets of Jacobi variables for internal motion of the three-body
system. Complete definitions of the hyperspherical variables and hyperspherical
harmonics can be found e.g.\ in Ref.\ \cite{gri03b}.
The dependence on magnetic quantum numbers is
\[
g^{JLSS_x}_{MM_LM_1M_2M_3}=\sum_{M_SM_x}
C^{JM}_{LM_L SM_S} C^{SM_S}_{S_xM_x S_3M_3} C^{S_xM_x}_{S_1M_1 S_2M_2}\, .
\]
Following the same steps as in the two-body case we define scattering amplitude
and decompose it over hyperspherical harmonics. Asymptotic form of the
three-body
WF (center of mass motion is omitted) is given by
\begin{equation}
\Psi_{3}(\rho\rightarrow\infty)=\Psi_{3}^{pw}+
\frac{\exp[i\varkappa\rho]}{\rho^{5/2}}  \,
f_{M_1M_2M_3}(\Omega_{\rho},\Omega_{\varkappa})
\label{eq:3b-ass}
\end{equation}
where
\begin{eqnarray}
f_{M_1M_2M_3}(\Omega_{\rho},\Omega_{\varkappa}) = \sum_{JM} \, \sum_{LSS_x} \,
\sum_{M_LM_S} \, C^{JM}_{LM_LSM_S}\nonumber \\
\times \;
f^{\,J\,L\,S\,S_x}_{MM_LM_1M_2M_3}(\Omega_{\rho},\Omega_{\varkappa})
\; X_{SS_xM_S}\, .
\nonumber
\end{eqnarray}
So, $3\rightarrow 3$ scattering amplitude can be written as
\begin{eqnarray}
f^{\,J\,L\,S\,S_x}_{MM_LM_1M_2M_3}(\Omega_{\rho},\Omega_{\varkappa})=
\exp[-i\pi/4] \; \left(2\pi/\varkappa \right)^{5/2}
\nonumber \\
\times \; \sum_{Kl_xl_y,K^{\prime}\gamma^{\prime}} \,
\bigl( \delta_{K\gamma}^{K^{\prime}\gamma^{\prime}}-
S_{K\gamma}^{K^{\prime}\gamma^{\prime}} \bigr) \;
\mathcal{I}^{LM_L}_{Kl_xl_y}(\Omega_{\rho})
\nonumber \\
\times \;  \sum_{M_L'} \,
\mathcal{I}_{K^{\prime}l_x'l_y'}^{L'M_L'\ast}(\Omega_{\varkappa})  \;
g^{JL'S'S_x'}_{MM_L'M_1M_2M_3} \, ,
\end{eqnarray}
where $\gamma=\{Ll_xl_ySS_x\}$.
In these equations angles $\Omega_{\varkappa}$ point to the direction  on the
hypersphere, where the particles have come from [the directions defined by
momenta in the plane wave (\ref{eq:pw3})]. The angles $\Omega_{\rho}$ in the
asymptotic expression (\ref{eq:3b-ass})
define the vectors of particles and the energy distribution after collision. The
cross section of $3\rightarrow3$ scattering can be written as
\[
\frac{d\sigma(\Omega_{\varkappa})}{d\Omega_{\rho}}= \!
\sum_{SS_xM_S}  \Bigl| \sum_{LM_L} C^{JM}_{LM_LSM_S}
f^{\,J\,L\,S\,S_x}_{MM_LM_1M_2M_3}
(\Omega_{\rho},\Omega_{\varkappa}) \Bigl|^{2} ,
\]
and after integration over the angle $\Omega_{\rho}$ of the outgoing
particles, summation over the projections of final total spin $M$ and averaging
over the projections of spins of the incoming clusters $M_i$
\begin{eqnarray}
\sigma(\Omega_{\varkappa})=\left( 2\pi/\varkappa \right)^{5} \; G^J_{S_1S_2S_3}
\,
\sum_{K\gamma,K^{\prime}\gamma^{\prime}} \delta^{S'}_{S} \, \delta^{S_x'}_{S_x}
\, \delta^{L'}_{L}\nonumber \\
\times  \; (2L+1)^{-1} \sum_{M_L} \;
\mathcal{I}^{LM_L}_{Kl_xl_y} (\Omega_{\varkappa}) \;
\mathcal{I}^{L'M_L\ast}_{K'l_x'l_y'}(\Omega_{\varkappa})
\nonumber \\
\times \, \sum_{K^{\prime\prime}\gamma^{\prime\prime}}
\bigl( \delta_{K^{\prime\prime}\gamma^{\prime\prime}}^{K\gamma}-
S_{K^{\prime\prime}\gamma^{\prime\prime}}^{K\gamma} \bigr)^{\dagger}
\bigl(\delta_{K^{\prime\prime}\gamma^{\prime\prime}}^{K^{\prime}\gamma
^{\prime}}-S_{K^{\prime\prime}\gamma^{\prime\prime}}^{K^{\prime}\gamma
^{\prime}}\bigr) \, .
\end{eqnarray}
Coefficients $G^J_{S_1S_2S_3}$ are combinatorial factors
\[
G^J_{S_1S_2S_3}=\frac{2J+1}{(2S_1+1)(2S_2+1)(2S_3+1)} \, .
\]
The astrophysical production rate is given by
\[
\left \langle \sigma_{2p,2p} v\right \rangle' =
2 \int\frac{\varkappa}{m} \;
\sigma(\Omega_{\varkappa}) \, w(k_{1} k_{2} k_{3}) \;
d^{3}k_{1}\,d^{3}k_{2}\,d^{3}k_{3} \, ,
\]
where the Boltzmann distribution for 3 particles is
\[
w(k_{1}k_{2}k_{3})=\frac{\exp[  -(E_{1}+E_{2}+E_{3})/kT]}
{(2\pi mkT)^{9/2} \, (A_{1}A_{2}A_{3})^{3/2}} \, .
\]
After transformation to hyperspherical variables
\[
d^{3}k_{1}d^{3}k_{2}d^{3}k_{3}\rightarrow
\left( \frac{A_{1}A_{2}A_{3}}{A_{1}+A_{2}+A_{3}} \right) ^{3/2}\,d^{3}k_{cm}\,
d\Omega_{\varkappa}\,\varkappa^{5}d\varkappa \, ,
\]
and after $d^{3}k_{cm}$ and $d\Omega_{\varkappa}$ integration (we should remind
here
that angles $\Omega_{\varkappa}$ point to the directions of the incoming
particles defined by vectors ${\bf k}_i$)
\begin{eqnarray}
\left \langle \sigma _{2p,2p}v \right \rangle' =
\frac{(2\pi)^{6}}{\pi(2\pi m k T)^{3}} \,
G^J_{S_1S_2S_3} \int \exp \left[-\frac{\varkappa^{2}}{2mkT}\right]
\,\frac{\varkappa}{m}
\nonumber \\
\times
\sum_{K\gamma,K^{\prime}\gamma^{\prime}}  \bigl(\delta_{K\gamma}^{K^{\prime}
\gamma^{\prime}}-S_{K\gamma}^{K^{\prime}\gamma^{\prime}}\bigr)^{\dagger}
\bigl(\delta_{K\gamma}^{K^{\prime}\gamma^{\prime}}-
S_{K\gamma}^{K^{\prime}\gamma^{\prime}}\bigr) \,
 d\varkappa \, . \quad
\end{eqnarray}
To get the inelastic part of the cross section for the sufficiently narrow
resonance we should replace \cite{baz77,fadd}
\begin{eqnarray}
\sum_{K\gamma,K^{\prime}\gamma^{\prime}}\bigl(\delta_{K\gamma}^{K^{\prime}
\gamma^{\prime}}-S_{K\gamma}^{K^{\prime}\gamma^{\prime}}\bigr)^{\dagger}
\bigl(\delta_{K\gamma}^{K^{\prime}\gamma}-S_{K\gamma}^{K^{\prime}\gamma^{\prime}
}\bigr) \nonumber \\
\rightarrow \frac{\Gamma_{2p}\Gamma_{\gamma}}
{(E-E_{3R})^{2}+\Gamma_{3R}^{2}/4} \,.
\end{eqnarray}
For the two-proton capture on the nucleus $J_I$ to the final state $J_F$
(assuming the small width of the three-body resonance) we obtain
\begin{equation}
\left \langle \sigma_{2p,\gamma}v\right \rangle' =
\left(  \frac{2\pi}{mkT}\right)^{3} \!\!
\frac{2J_{F}+1}{2(2J_{I}+1)}
\exp\left[  -\frac{E_{3R}}{kT}\right]
\frac{\Gamma_{2p}\Gamma_{\gamma}}{\Gamma_{3R}}.
\label{eq:prod-rate-direct-prime}
\end{equation}
The production rate for two-proton capture should be multiplied by squared
density $\rho^{2}$ to provide the abundance, see Eq.\ (\ref{eq:abund}).
Eq.\ (\ref{eq:prod-rate-direct-prime}) is written in scaled Jacobi
variables. The density in these variables can be expressed via density in
the ordinary space as
\[
\rho_{\text{scaled}}^{2}=\rho^{2}\;
\left(  \frac{A_{1}+A_{2}+A_{3}}{A_{1}A_{2}A_{3}}\right)^{3/2} .
\]
The expression for production rate which can be used with expression for
density in normal space (which is indicated by the absence of the prime symbol)
is therefore
\begin{eqnarray}
\left\langle \sigma_{2p,\gamma}v\right\rangle & = &
\left( \frac{A_{1}+A_{2} + A_{3}}{A_{1}A_{2}A_{3}}\right)^{3/2} \!\!
\frac{2J_{F}+1}{2(2J_{I}+1)} \left(\frac{2\pi}{m k T}\right)^{3} \nonumber \\
&\times & \, \exp\left[ -E_{3R}/kT \right] \;
\Gamma_{2p}\Gamma_{\gamma}/\Gamma_{3R} \, .
\label{eq:prod-rate-direct}
\end{eqnarray}
Eq.\ (\ref{eq:prod-rate-direct}) is absolutely the same as Eq.\
(\ref{eq:prod-rate-seq-1}) except for the dependence on the decay width to the
three-body continuum $\Gamma_{2p}$ instead of decay widths to the resonant
states in $A+1$ system $\Gamma_{p,i}^{\prime}$. We can draw the following
conclusions here:

\noindent (i) Eq. (\ref{eq:prod-rate-direct}) is obtained in a very general
assumptions about existence of the asymptotic (\ref{eq:3b-ass}) and analytical
properties of the $3 \rightarrow 3$ scattering S-matrix.
We also use the fact that most of the states of interest are narrow. No
other assumptions is made (e.g.\ in sequential formalism there is assumption
about existence of the specific decay path). So, the direct capture (unlike
sequential capture) is always possible.

\noindent (ii) It is clear that Eqs.\ (\ref{eq:prod-rate-seq-1}) and
(\ref{eq:prod-rate-direct}) supplement each other and the total reaction rate is
\begin{eqnarray}
\left\langle \sigma_{2p,\gamma}v\right\rangle +
\left\langle \sigma_{pp,\gamma}v\right\rangle =
\left( \frac{A_{1}+A_{2} + A_{3}}{A_{1}A_{2}A_{3}}\right)^{3/2} \!\!
\frac{2J_{F}+1}{2(2J_{I}+1)}  \nonumber \\
\times \,\left(\frac{2\pi}{m k T}\right)^{3}  \exp\left[ -\frac{E_{3R}}{k T}
\right] \frac{\Gamma_{2p}+\textstyle \sum_i\Gamma_{p,i}^{\prime}}{\Gamma_{3R}}
\, \Gamma_{\gamma} \, . \quad
\label{eq:prod-rate-total}
\end{eqnarray}

\noindent (iii) In the most likely situation
$\Gamma_{2p}+\textstyle \sum_i\Gamma_{p,i}^{\prime} \gg \Gamma_{\gamma}$ (and
hence $\Gamma_{3R}=\Gamma_{2p}+\textstyle \sum_i\Gamma_{p,i}^{\prime}$) the
total reaction rate depends {\em only} on gamma width of the three-body
resonance. However, it is possible that the direct two-proton emission is the
only nuclear decay branch for the state. The width $\Gamma_{2p}$ could be very
small (smaller than the gamma width) in a relatively broad range of the
three-body decay energies \cite{gri03b}. In that case the reaction rate depends
{\em only} on $\Gamma_{2p}$.

\noindent (iv) Eq.\ (\ref{eq:prod-rate-total}) gives the formula for reaction
rate, which is the same as one, known for a long time (see e.g.\ Eq.\ (20) in
Ref.\ \cite{fow67}), which was obtained by much easier means (namely, a complete
thermal equilibrium and a detailed balance) than in this work. The result of our
derivation here is clearer understanding of the fact that this formula already
correctly and completely include both sequential
capture and direct capture reactions.

So, we see that for the resonant part of the reaction rate the sequential
formalism treatment is overcomplicated and incomplete. This is not a great issue
in most cases, but there are situations, where it becomes important. The impact
of the formalism on the rates of reactions of astrophysical interest is
discussed in Sec.\ \ref{sec:disc}.


\subsection{Formal questions}

\label{sec:formal}

The derivations of the reaction rates above in Sections \ref{sec:seq-cap} and
\ref{sec:dir-capt} are quite schematic. They basically rely on assumptions about
existence of definite asymptotics of the three-body problem, which could be not
evident. They require if not a proof then at least some discussion.

For sequential formalism we need that there exists a long-living resonance state
in the $X$ Jacobi subsystem (at energy $E_x=k_x/(2M_x)$ and with width
$\Gamma_x$). Then the asymptotic implied in derivations of Sec.\
\ref{sec:seq-cap} is
\begin{equation}
\Psi_3(\{X,Y\} \rightarrow \infty) =
\Psi_3^{pw} + \frac{e^{ik_xX}}{X}f(\hat{k}_x)
\frac{e^{ik_yY}}{Y}f(\hat{k}_y) \, .
%
\label{eq:seq-ass}
\end{equation}
For the direct capture the assumed asymptotic is
\begin{equation}
\Psi_3(\rho \rightarrow \infty) = \Psi_3^{pw} +
\frac{e^{i\varkappa\rho}}{ \rho^{5/2}} \, f(\Omega_\rho,\Omega_\varkappa)\, ,
\label{eq:dir-ass}
\end{equation}
where $k_x^2/(2M_x)+k_y^2/(2M_y)=\varkappa^2/(2m)=E_{3R}$.

The expressions (\ref{eq:seq-ass}) and (\ref{eq:dir-ass}) correspond to neutral
particles, while we are speaking about nuclei $Z=8-20$ capturing protons.
The typical densities for X-ray bursts and processes in novae are
$10^{3}-10^{6}$ g/ccm \cite{wie99}. For such densities the average distances
between protons are about $6\times 10^2-2\times 10^4$ fm and for characteristic
temperatures the Debye screening radii are $3\times 10^{3}-5\times 10^5$ fm.
Beyond these radii we have a formal right to use the forms (\ref{eq:seq-ass})
and
(\ref{eq:dir-ass}).

It should be understood that in a very formal sense the asymptotic
(\ref{eq:seq-ass}) is not valid. In the limit of infinite distance a long-living
two-body state finally decays
%
%
and the asymptotic (\ref{eq:seq-ass}) should be replaced by (\ref{eq:dir-ass}).
However, from practical side the separation of asymptotics (\ref{eq:seq-ass})
and (\ref{eq:dir-ass}) is reasonable. A nice example (also relevant to the
further discussion) of the coexistence of the three-body and binary asymptotics
is the decay of $^9$Be $5/2^-$ state at 2.429 MeV. The branchings to the
three-body channel $Br(3)$ and binary channel $Br(2)$ are comparable
($Br(3)\sim 0.93-0.95$ and $Br(2)\sim 0.07-0.05$ \cite{ajz88}),
which is experimentally well observed (see, for example, Ref.\ \cite{nym90}). In
the case of binary decay the average flight distance of the $^8$Be g.s.\
resonance
($\Gamma_x=6.8$ eV) is around $10^6$ fm. Thus it is clear that for some
practical purposes the assumption of Eq.\ (\ref{eq:seq-ass}) is justified. It
is also clear that the broader is the resonance in the $X$ subsystem, the
faster a transition from Eq.\ (\ref{eq:seq-ass}) to Eq.\ (\ref{eq:dir-ass})
happens. For example, for the width of the intermediate resonance $\Gamma_x$
around 100 keV (and typical $E_{3R}=1$ MeV) the average flight distance of this
resonance is around 100 fm. This is much smaller than the typical distance
between protons in the stellar media and then usage of Eq.\ (\ref{eq:seq-ass})
and chemical balance description by the Eqs.\ (\ref{eq:sys-eq}) loose sense.

The asymptotics (\ref{eq:seq-ass}) and (\ref{eq:dir-ass}) are located in the
same space and could have the same quantum numbers.
However, we can can speak about
orthogonality of these asymptotics in definite sense and then treat currents
associated with them independent. Only this assumption makes possible the
separate treatment of sequential and direct decay channels in Sections
\ref{sec:seq-cap} and \ref{sec:dir-capt}. The asymptotic (\ref{eq:seq-ass}) is
typically localized in a very small part of the phase space. Formally this
corresponds to the fact that the hyperspherical series for binary channel (at
given hyperradius) is very long (with many significant terms), while for
asymptotic (\ref{eq:dir-ass}) we can expect that only the lowest hyperspherical
harmonics in
the decomposition are significant. Thus the sequential and direct channels are
practically orthogonal on the hypersphere of a large radius. With increase of
the radius this ``orthogonality'' (called asymptotic orthogonality) becomes
better, until the effect of the $X$ subsystem  decay becomes important. Again,
on the example of the $5/2^-$ state of $^9$Be, the level of overlap between the
direct and sequential decay channels in the momentum space can be estimated as
\[
Br(3)\,Br(2)\,\frac{\Gamma_{^8\text{Be}}(0^+)}{E_{3R}(5/2^-)} =
0.06\,\frac{6.8 \text{ eV}}{ 764.1 \text{ keV}} \sim
5\times 10^{-7} ,
\]
which is clearly a very small value. When there is no such a reliable separation
between channels (\ref{eq:seq-ass}) and (\ref{eq:dir-ass}) any more (the
intermediate resonances are too broad) we have a formal right to speak only
about asymptotic (\ref{eq:dir-ass}).

To finalize this discussion, the derivations in Sections \ref{sec:seq-cap} and
\ref{sec:dir-capt} were done as if only one type of the asymptotic exists. From
formal point of view only asymptotic of Eq.\ (\ref{eq:dir-ass}) exists. For
practical purposes either (i) only asymptotic of Eq.\ (\ref{eq:dir-ass}) exists
or (ii) both asymptotics Eqs.\ (\ref{eq:seq-ass}) and (\ref{eq:dir-ass}) are 
present simultaneously. The asymptotic (\ref{eq:dir-ass})
exists in the three-body problem unconditionally, while the existence of
(\ref{eq:seq-ass}) is subject to availability of {\em sufficiently narrow}
intermediate resonances for the decay. In the case (ii), the regions of the
domination of each asymptotic are separated by a complicated surface in the
phase space. Some discussion of the relevant questions can be found, for
example, in Ref.\ \cite{fadd}. So, the phase space integration in Sections
\ref{sec:seq-cap} and \ref{sec:dir-capt} should have been done not over all the
space, but over regions of validity for each type of asymptotic. This is clear,
however, that this imperfection does not influence the final result. The reasons
are that (i) the contribution of the asymptotic of the selected kind in the
phase
space outside the region of its domination is typically negligible and (ii)
anyhow we are interested in the contribution of the both kinds of asymptotics
simultaneously.


\section{Discussion}

\label{sec:disc}


\subsection{$^{15}$O($2p$,$\gamma$)$^{17}$Ne reaction}

\begin{figure}
\includegraphics[width=0.47\textwidth]{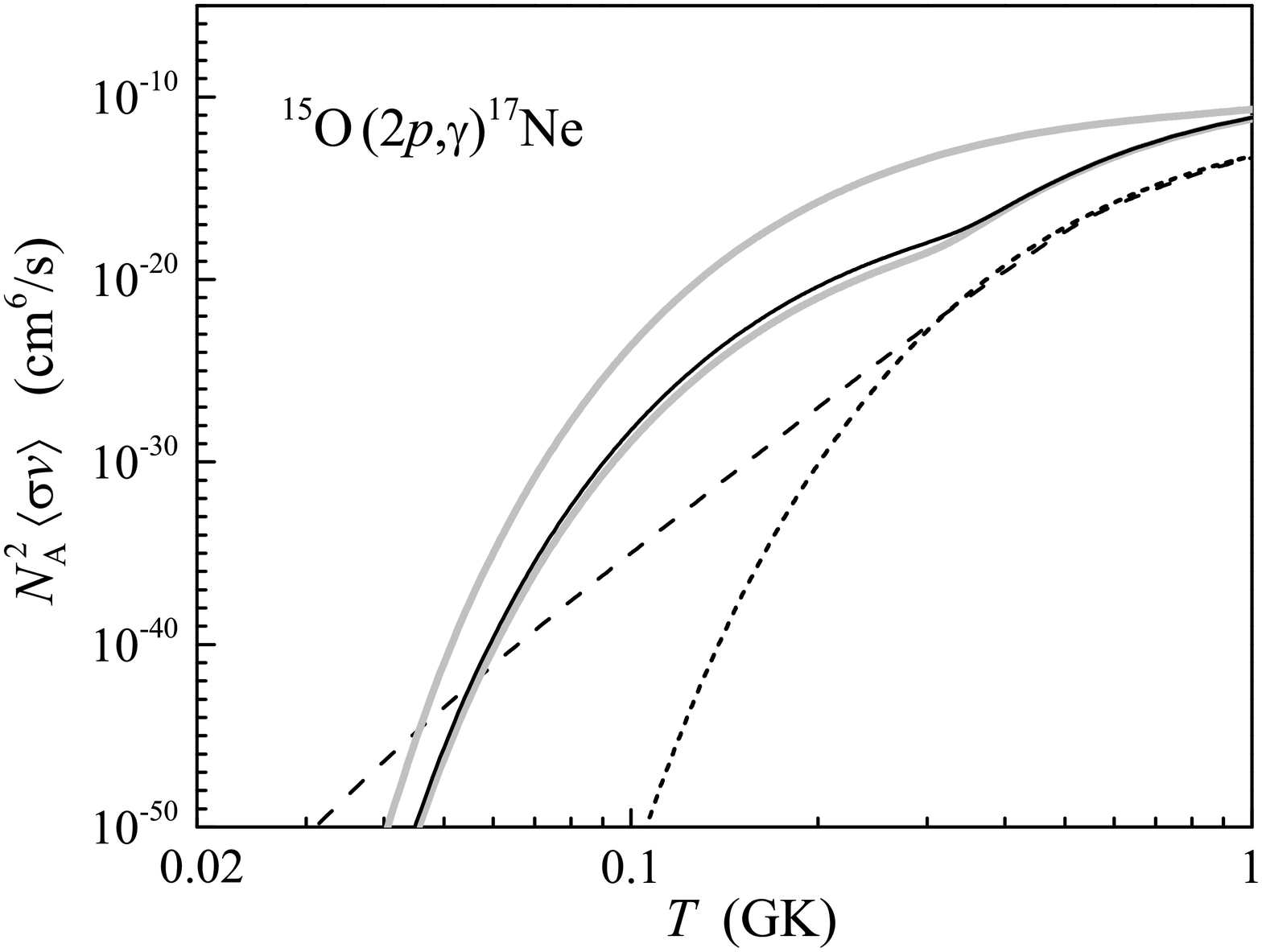}
\caption{Reaction rate for $^{15}$O($2p$,$\gamma$)$^{17}$Ne reaction. Solid
curve shows calculations of this paper. Gray curves indicate boundaries due to
uncertainties in the input (e.g.\ the upper gray curve is obtained with
parameter set $\Gamma_{\gamma}^{\max}$ from Table \protect\ref{tab:17ne-par}).
Dashed and dotted curves show full result from Ref.\ \cite{gor95} and resonance
contribution to it respectively.}
\label{fig:17ne-rate}
\end{figure}

The results of rate calculations for this reaction are shown in Fig.\
\ref{fig:17ne-rate} and in  Table \ref{tab:17ne-rate}. They differ
significantly from the results of Ref.\ \cite{gor95} (shown in Fig.\
\ref{fig:17ne-rate} by dashed and dotted curves). For the temperature range of
astrophysical interest ($\sim 0.3-3$ GK, see \cite{wie99}, for example) the
expected increase of the rate, compared to Ref.\ \cite{gor95}, is up to 4 orders 
of the magnitude, while maximal possible increase is up to 9 orders of the
magnitude.

The reasons of difference are evident from Table \ref{tab:17ne-par}.

\noindent (i) The level scheme of $^{17}$Ne has been somewhat updated (see
e.g.\ Ref.\ \cite{gui98}) since the the work \cite{gor95} had been written.

\noindent (ii) The use of Eq.\ (\ref{eq:prod-rate-direct}) includes the first
$3/2^-$ excited state of $^{17}$Ne into treatment (it was omitted in Ref.\
\cite{gor95}, as there is no sequential capture path to this state). The
important difference of the situation with this state from the others is that
gamma width of this state is known to be much larger than the $2p$ width and
the reaction rate Eq.\ (\ref{eq:prod-rate-total}) is entirely defined by the
$2p$ width for the simultaneous two-proton emission. At the moment there exist
two theoretical calculations of this width $\Gamma_{2p}=4.1 \times 10^{-16}$
MeV \cite{gri03}, $\Gamma_{2p}=3.6 \times 10^{-12}$ MeV \cite{gar04}, and a
quite relaxed experimental lower lifetime limit of $\tau >
26$ ps \cite{chr02} (which corresponds to the width $\Gamma_{2p}<2.5 \times
10^{-11}$ MeV). Values $\Gamma_{2p}$ from Refs.\ \cite{gri03} and
\cite{chr02} are used to estimate, respectively, the lower and the upper
boundaries for the band of expected values of the rate (see Fig.\
\ref{fig:17ne-rate}).
The resonance contribution of this state is dominating the rate
in the temperature range $0.05-0.35$ GK if we take theoretical $2p$ width from
Ref.\ \cite{gri03}, and up to 1.2 GK if we consider the experimental limit.

\begin{table}
\caption{Reaction rates multiplied by $N^2_A$ (in cm$^6$/s) for
$^{15}$O($2p$,$\gamma$)$^{17}$Ne reaction.}
\vspace{2mm}
\begin{ruledtabular}
\begin{tabular}[c]{cccc}
$T$ (GK) & Ref.\ \cite{gor95}  & This work & This work upper\\
\hline
 0.3  & $2.9\times 10^{-23}$ & $4.9\times 10^{-19}$ & $2.9\times 10^{-14}$ \\
 0.5  & $6.0\times 10^{-18}$ & $2.1\times 10^{-15}$ & $1.3\times 10^{-12}$ \\
 0.6  & $1.2\times 10^{-16}$ & $2.8\times 10^{-14}$ & $2.8\times 10^{-12}$ \\
 0.8  & $5.6\times 10^{-15}$ & $6.3\times 10^{-13}$ & $6.9\times 10^{-12}$ \\
 1.0  & $5.0\times 10^{-14}$ & $3.5\times 10^{-12}$ & $1.3\times 10^{-11}$ \\
 1.5  & $1.1\times 10^{-12}$ & $2.5\times 10^{-11}$ & $3.8\times 10^{-11}$ \\
 2.0  & $6.0\times 10^{-12}$ & $5.1\times 10^{-11}$ & $6.8\times 10^{-11}$ \\
 3.0  & $4.3\times 10^{-11}$ & $7.6\times 10^{-11}$ & $1.1\times 10^{-10}$ \\
 5.0  & $2.5\times 10^{-10}$ & $7.3\times 10^{-11}$ & $2.3\times 10^{-10}$ \\
\end{tabular}
\end{ruledtabular}
\label{tab:17ne-rate}
\end{table}

\noindent (iii) In paper \cite{gor95} the gamma widths for $^{17}$Ne were taken
from studied transitions in the isobaric mirror partner $^{17}$N. Recently the
decay of the first excited states of $^{17}$Ne (3/2$^{-}$, 5/2$^{-}$) has been
studied via intermediate energy Coulomb excitation of a radioactive $^{17}$Ne
beam on a $^{197}$Au target \cite{chr02}. In this paper the transition matrix
elements $B($E2,$ 1/2^- \! \rightarrow 3/2^-)$ and $B($E2,$ 1/2^- \!\rightarrow
5/2^-)$ have been deduced. We use the deduced $B($E2$)$  value from Ref.\
\cite{chr02} to calculate the gamma width of the $5/2^-$ state. The result is
shown in Table \ref{tab:17ne-par}. This width appears to be about 30 times
larger that the corresponding width of the mirror state in $^{17}$N. This is,
probably, connected to the fact that at the proton-rich side the number of the
protons contributing gamma transitions is larger and these protons are situated
at larger distances compared with tightly bound protons in $^{17}$N. This
situation is also expected for the other states in $^{17}$Ne (compared to the
states in $^{17}$N), which is reflected by an order of the magnitude increase
of  the other widths for estimates of the upper limits (column
$\Gamma_{\gamma}^{\max}$ in Table \ref{tab:17ne-par}).

\begin{table}
\caption{Resonance parameters of $^{17}$Ne states used in the
$^{15}$O($2p$,$\gamma$)$^{17}$Ne reaction calculations. The column
$\Gamma_{\gamma}^{\max}$ shows the upper experimental limit for
$\Gamma_{\gamma}$ values, and $\Gamma_{\gamma}^{\min}$ the values
used for lower limit estimates. The listed set of states is sufficient for
rate calculations up to 5 GK. The widths and branchings, which are not specially
discussed are from \cite{ajz88}.}
\vspace{2mm}
\begin{ruledtabular}
\begin{tabular}[c]{cccccc}
State  & \multicolumn{2}{c}{Ref.\ \cite{gor95}} & \multicolumn{3}{c}{This work}
 \\
$J^{\pi}$ & $E$ (keV) & $\Gamma_{\gamma}$ (eV) & $E$ (keV) &
$\Gamma^{\min}_{\gamma}$ (eV) & $\Gamma_{\gamma}^{\max}$ (eV)  \\
\hline
 $3/2^-$ &       &                     & 1288 & \footnotemark[1]$4.1\times
10^{-10}$  & \footnotemark[2]$2.5\times 10^{-5}$    \\
 $5/2^-$ & 1907 & $6.0\times 10^{-5}$ & 1764 & \footnotemark[3]$1.7\times
10^{-3}$ & \footnotemark[3]$2.0\times 10^{-3}$   \\
 $1/2^+$ & 1850 & $1.6\times 10^{-5}$ & 1908 & $1.6\times 10^{-5}$ &
\footnotemark[6]$2.1\times
10^{-4}$   \\
 $5/2^+$ & 2526 & $2.0\times 10^{-5}$ & 2651 & \footnotemark[4]$9.0\times
10^{-6}$   &  \footnotemark[6]$9.9\times 10^{-5}$   \\
 $3/2^-$ & 3204 &  $0.022$ & 3204 &  \footnotemark[5]0.019  &
\footnotemark[6]0.19 \\
\end{tabular}
\end{ruledtabular}
\footnotetext[1]{This is a $2p$ width (see Eq.\ (\ref{eq:prod-rate-total}): the
gamma width is dominating the decay of this state). This value is calculated
theoretically in Ref.\ \cite{gri03}.}
\footnotetext[2]{This is a $2p$ width. This experimental limit on $2p$ width is
found in Ref.\ \cite{chr02}.}
\footnotetext[3]{This value is calculated from $B($E2$)=124(18)$ e$^2$fm$^4$
given in Ref.\ \cite{chr02}.}
\footnotetext[4]{This value is partial width ($45\%$ branching) into
the $1/2^-$ ground and the first excited $3/2^-$ states of $^{17}$Ne. The gamma
transition to $1/2^+$ state returns the system into $2p$ continuum.}
\footnotetext[5]{This value is partial width ($88\%$ branching) into
the ground and first excited states of $^{17}$Ne.}
\footnotetext[6]{These values are assumed (in analogy with more than order of
magnitude increase for $5/2^-$ state from column 3 to column 5).}
\label{tab:17ne-par}
\end{table}


\subsection{$^{18}$Ne($2p$,$\gamma$)$^{20}$Mg reaction}

For the $^{18}$Ne($2p$,$\gamma$)$^{20}$Mg reaction there are no three-body
states which were not taken into account in Ref.\ \cite{gor95}, so, no
significant update of the rate is expected here. However, the level scheme and
gamma widths are not known experimentally for this nucleus and this should be
reflected in the rate calculations

\begin{table}
\caption{Resonance parameters of $^{20}$Mg states used in the
$^{18}$Ne($2p$,$\gamma$)$^{20}$Mg reaction calculations. }
\vspace{2mm}
\begin{ruledtabular}
\begin{tabular}[c]{ccccc}
State  & \multicolumn{2}{c}{This work lower} & \multicolumn{2}{c}{This work
upper}
 \\
$J^{\pi}$ & $E$ (keV) & $\Gamma_{\gamma}$ (eV) & $E$ (keV) & $\Gamma_{\gamma}$
(eV)   \\
\hline
 $4^+_1$ & 3570 & $2.1\times 10^{-4}$  & 3451 & $5.6\times 10^{-4}$ \\
 $2^+_2$ & 4072 & $1.3\times 10^{-3}$  & 3857 & $8.9\times 10^{-2}$ \\
 $0^+_2$ & 4456 & $1.4\times 10^{-3}$  & 4317 & $1.4\times 10^{-3}$ \\
 $4^+_2$ & 4850 & $2.6\times 10^{-3}$  & 4699 & $2.6\times 10^{-3}$ \\
 $2^+_3$ & 5234 & $2.9\times 10^{-1}$  & 4978 & $2.9\times 10^{-1}$ \\
\end{tabular}
\end{ruledtabular}
\label{tab:20mg-par}
\end{table}

In the cases of $2p$ capture into $^{17}$Ne and $^{40}$Ti the gamma widths from
mirror isobaric partners were used in Ref.\ \cite{gor95}. In contrast, for
capture into $^{20}$Mg the systematics values were utilized. The theoretical
$B$(E2) values for some low-lying states in $^{20}$Mg have been calculated
recently in Ref.\ \cite{des98}. The $B$(E2,$4^+_1 \!\rightarrow 2^+_1$) was
found to be 28.2 or 11.6 $e^2$fm$^4$ (for V2 and MN forces respectively) and
$B$(E2,$2^+_2 \!\rightarrow 0^+_1$)  was found to be  2.9 or 1.9 $e^2$fm$^4$.
These reduced probabilities give gamma widths $5.6 \times 10^{-4}$ or $2.3
\times 10^{-4}$ eV for $4^+_1$ state (which is comparable to value $2.1 \times
10^{-4}$ eV used in \cite{gor95}) and $2.0 \times 10^{-3}$ or $1.3 \times
10^{-3}$ eV for $2^+_2$ state (which is significantly less than $8.9 \times
10^{-2}$ eV used in \cite{gor95}). We combine the largest and the lowest gamma
widths from Refs.\ \cite{gor95} and \cite{des98} to estimate the upper and the
lower boundaries for the rate (see Table \ref{tab:20mg-par}). To incorporate in
this estimate the sensitivity to the level scheme we also use for the lower
estimate the energies of the states from $^{20}$O. The distance between levels
here is expected to be somewhat larger than in $^{20}$Mg \cite{gor95} and the
reaction rate thus should further decrease.

\begin{table}
\caption{Reaction rates multiplied by $N^2_A$ (in cm$^6$/s) for
$^{18}$Ne($2p$,$\gamma$)$^{20}$Mg reaction.}
\vspace{2mm}
\begin{ruledtabular}
\begin{tabular}[c]{cccc}
$T$ (GK) & Ref.\ \cite{gor95}  & This work lower & This work upper \\
\hline
 0.3  & $4.4\times 10^{-21}$ & $9.3\times 10^{-28}$  & $2.5\times 10^{-25}$  \\
 0.5  & $3.3\times 10^{-17}$ & $4.3\times 10^{-20}$  & $1.8\times 10^{-18}$  \\
 0.6  & $4.8\times 10^{-16}$ & $3.0\times 10^{-18}$  & $8.3\times 10^{-17}$  \\
 0.8  & $1.8\times 10^{-14}$ & $5.1\times 10^{-16}$  & $9.5\times 10^{-15}$  \\
 1.0  & $2.4\times 10^{-13}$ & $9.6\times 10^{-15}$  & $1.8\times 10^{-13}$  \\
 1.5  & $6.8\times 10^{-12}$ & $3.7\times 10^{-13}$  & $1.1\times 10^{-11}$  \\
 2.0  & $9.7\times 10^{-11}$ & $2.0\times 10^{-12}$  & $8.0\times 10^{-11}$  \\
 3.0  & $4.8\times 10^{-10}$ & $1.5\times 10^{-11}$  & $4.4\times 10^{-10}$  \\
 5.0  & $2.4\times 10^{-9}$  & $1.4\times 10^{-10}$  & $1.2\times 10^{-9}$  \\
\end{tabular}
\end{ruledtabular}
\label{tab:20mg-rate}
\end{table}

The results of calculations are shown in Table \ref{tab:20mg-rate}. The upper
boundary in our calculations is in a good agreement with results of \cite{gor95}
(factor of two) at $T \geq 0.8$ GK. It was shown in \cite{gor95} that below 0.8
GK the nonresonant contribution to the reaction rate dominates, which explains
the discrepancy in Table \ref{tab:20mg-rate} at low temperatures.


\subsection{$^{38}$Ca($2p$,$\gamma$)$^{40}$Ti reaction}

For the $^{38}$Ca($2p$,$\gamma$)$^{40}$Ti reaction one three-body state was
omitted in Ref.\ \cite{gor95}. According to the isobaric symmetry there should
be a $0^+_2$ state located at about 2.121 MeV excitation energy. The two-proton
separation energy used in Ref.\ \cite{gor95} is $S_{2p}=1.582$ MeV. Another
estimate (e.g.\ \cite{nndc}) is  $S_{2p}=1.370$ MeV. In the first case the $2p$
emission energy for $0^+_2$ state is 539 keV and (following Refs.\
\cite{gri03a,gri03b}) the two-proton width can be estimated as about $10^{-21}$
MeV. In the second case the $2p$ energy is 751 keV and the estimated two-proton
width is around $10^{-18}$ MeV. For other states we use parameters from Ref.\
\cite{gor95} (see Table \ref{tab:40ti-par}), which mainly come from the isobaric
mirror partner $^{40}$Ar. To estimate the upper boundary for the reaction rate
we increase the gamma widths of $4^+_1$ and $2^+_3$ states by an order of the
magnitude. As we have already discussed, one could expect a significant
increase of the gamma widths when we come to the proton-rich mirror partner. To
estimate the sensitivity to the level scheme (which is not known for $^{40}$Ti)
we use the smaller $2p$ separation energy $S_{2p}=1.370$ MeV for the estimate
of the lower boundary and $S_{2p}=1.582$ MeV for the upper boundary. Again, as
in the case of $^{20}$Mg, the increase of the state energy above $2p$ threshold
leads to decrease of the corresponding reaction rate. For that reason we use
the larger $2p$ width for estimate of the lower boundary (Table
\ref{tab:40ti-par}, line 1). The larger two proton width corresponds to the
case of larger energy of states above $2p$ separation threshold.

\begin{table}
\caption{Resonance parameters of $^{40}$Ti states used in the
$^{38}$Ca($2p$,$\gamma$)$^{40}$Ti reaction calculations. In the ``Type'' column
the type of width is specified, which defines the contribution of the state to
reaction rate. The ``Lower'' set of widths is used in the calculations with
$S_{2p}=1.370$ MeV, while the ``Upper'' set with $S_{2p}=1.582$ MeV.}
\vspace{2mm}
\begin{ruledtabular}
\begin{tabular}[c]{ccccc}
$J^{\pi}$ & $E$ (keV) & Type & ``Lower'' $\Gamma$ (eV)
& ``Upper'' $\Gamma$ (eV)  \\
\hline
 $0^+_2$ & 2121 & $\Gamma_{2p}$ & $10^{-12} $ & $10^{-15}$ \\
 $2^+_2$ & 2524 & $\Gamma_{p}$  & $1.0\times 10^{-5}$ & $1.0\times 10^{-5}$ \\
 $4^+_1$ & 2892 & $\Gamma_{\gamma}$ & $2.0\times 10^{-4}$ & $2.0\times 10^{-3}$
 \\
 $2^+_3$ & 3208 & $\Gamma_{\gamma}$ & $1.0\times 10^{-2}$ & $1.0\times 10^{-1}$
\\
\end{tabular}
\end{ruledtabular}
\label{tab:40ti-par}
\end{table}

\begin{figure}
\includegraphics[width=0.47\textwidth]{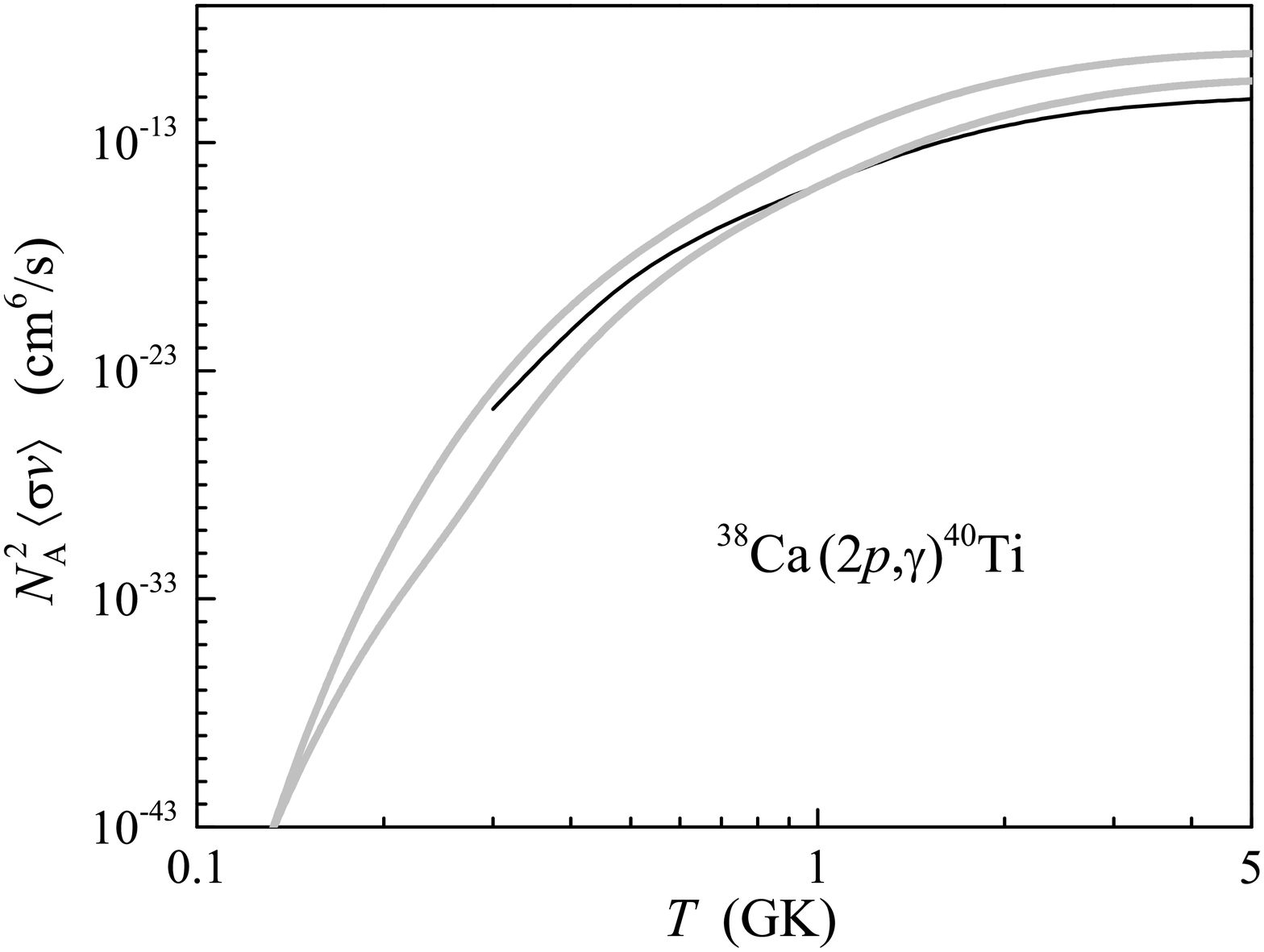}
\caption{Reaction rate for $^{38}$Ca($2p$,$\gamma$)$^{40}$Ti reaction. Solid
curve shows the result from \cite{gor95}. Gray curves indicate upper and lower
boundaries for our results (see Table \ref{tab:40ti-rate}) due to existing
uncertainties in the input.}
\label{fig:40ti-rate}
\end{figure}

The results of calculations for $^{38}$Ca($2p$,$\gamma$)$^{40}$Ti are given in
Fig.\ \ref{fig:40ti-rate} and Table \ref{tab:40ti-rate}. Our results are
somewhat larger (1--2 orders of magnitude) than results of \cite{gor95} for
temperatures $T>1$ GK. They more or less overlaps at lower temperatures. The
effect of inclusion of $0^+_2$ 2.121 MeV state can be seen in Fig.\
\ref{fig:40ti-rate}: the range between upper and lower boundaries shrinks at
$T<0.35$ GK. This happens because the contribution of the $0^+_2$ state is much
larger in the ``Lower'' parameter set, which otherwise provides a smaller
reaction
rate.

\begin{table}
\caption{Reaction rates multiplied by $N^2_A$ (in cm$^6$/s) for
$^{38}$Ca($2p$,$\gamma$)$^{40}$Ti reaction.}
\vspace{2mm}
\begin{ruledtabular}
\begin{tabular}[c]{cccc}
$T$ (GK) & Ref.\ \cite{gor95}  & This work lower & This work upper\\
\hline
 0.3  & $2.1\times 10^{-25}$ & $7.0\times 10^{-28}$ & $2.3\times 10^{-24}$ \\
 0.5  & $1.0\times 10^{-19}$ & $7.8\times 10^{-21}$ & $1.1\times 10^{-18}$ \\
 0.6  & $2.4\times 10^{-18}$ & $4.0\times 10^{-19}$ & $3.1\times 10^{-17}$ \\
 0.8  & $1.1\times 10^{-16}$ & $5.3\times 10^{-17}$ & $3.0\times 10^{-15}$ \\
 1.0  & $1.3\times 10^{-15}$ & $1.2\times 10^{-15}$ & $7.1\times 10^{-14}$ \\
 1.5  & $7.0\times 10^{-14}$ & $1.3\times 10^{-13}$ & $6.0\times 10^{-12}$ \\
 2.0  & $5.2\times 10^{-13}$ & $1.5\times 10^{-12}$ & $5.1\times 10^{-11}$ \\
 3.0  & $3.0\times 10^{-12}$ & $1.4\times 10^{-11}$ & $3.2\times 10^{-10}$ \\
 5.0  & $8.0\times 10^{-12}$ & $5.0\times 10^{-11}$ & $8.2\times 10^{-10}$ \\
\end{tabular}
\end{ruledtabular}
\label{tab:40ti-rate}
\end{table}

\subsection{$^4$He($n\alpha$,$\gamma$)$^{9}$Be reaction}

The stellar reaction rate for $^4$He($n\alpha$,$\gamma$)$^{9}$Be process has
been studied several times in the recent years
\cite{fow75,cau88,gor95a,efr98,ang99,buc01}. The results are in overall
agreement, except for the latest paper \cite{buc01}. In this work the rate is
obtained which is significantly higher (for temperatures $T>3$ GK) than the
rates in the previous studies.

\begin{figure}
\includegraphics[width=0.47\textwidth]{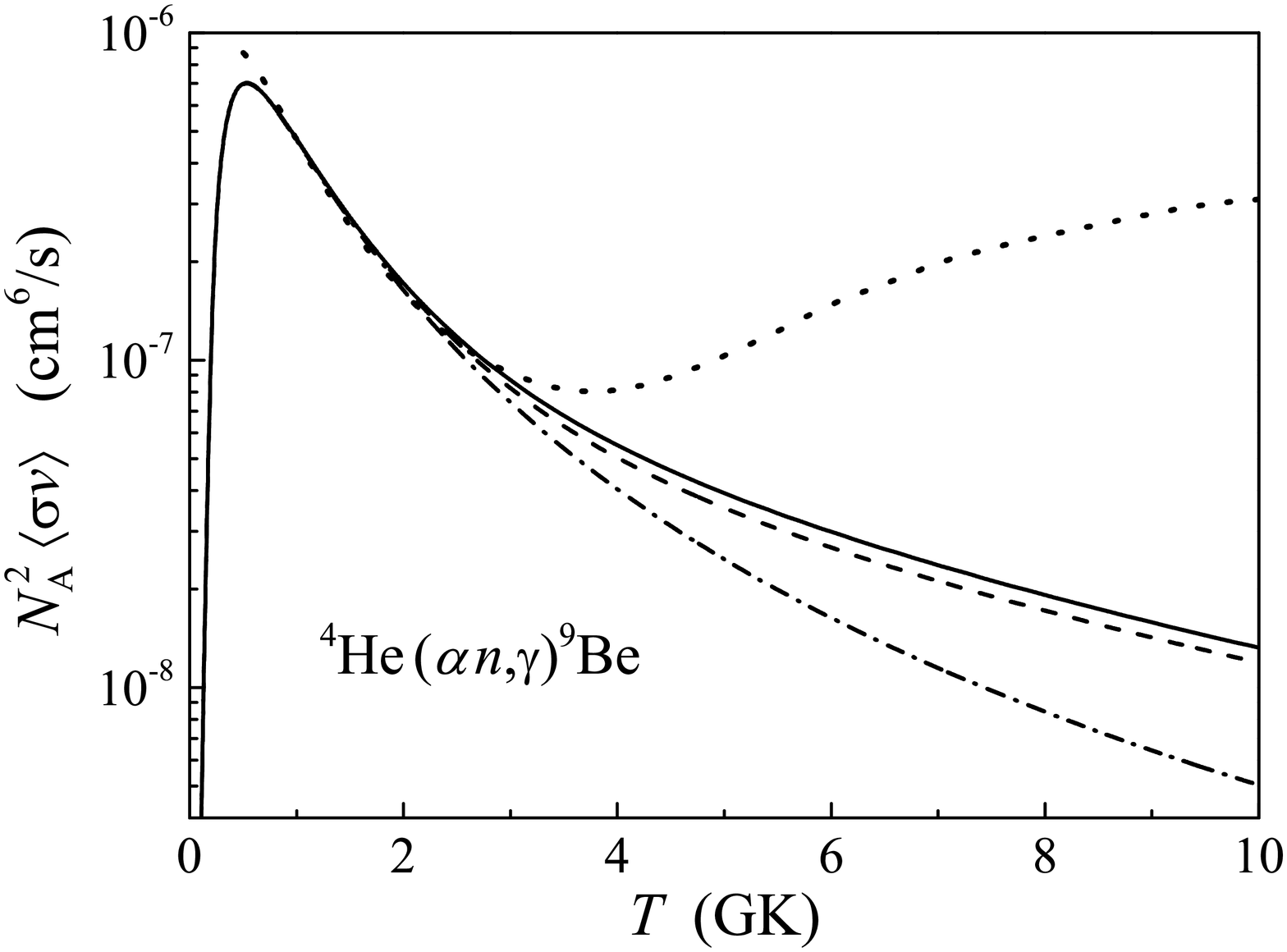}
\caption{Reaction rate for $^4$He($n\alpha$,$\gamma$)$^{9}$Be reaction. The
solid curve shows our result (see also Table \ref{tab:9be-rate}). The dashed
curve is the same, but without $5/2^-$ 2.429 MeV state contribution (this
coincides with result of Ref.\ \cite{ang99}). The dash-dotted curve is
contribution from the near threshold $1/2^+$ state. The dotted curve shows
calculations Ref.\ \cite{buc01}.}
\label{fig:9be-rate}
\end{figure}

In our studies here we have found that the sequential formalism underestimate
the reaction rate only if the width of the state for direct decay into
continuum is dominating (see Section \ref{sec:seq-cap}). The low-lying ($E \leq
3$ MeV) $^9$Be states typically have strong $^8$Be+$n$ decay branchings. Only
the $5/2^-$ 2.429 MeV state is an exception: the branching to the three-body
channel is $93-95\%$ \cite{ajz88}. The gamma width of this state is 0.091 eV
\cite{ajz88}. The results of our calculations are shown in Fig.\
\ref{fig:9be-rate} and Table \ref{tab:9be-rate}. In these calculations we use a
version of Eq.\ (\ref{eq:prod-rate-total}) without an assumption about narrow
widths of the resonances and the capture cross section is parametrized as in
Ref.\ \cite{ang99} (with exception that $5/2^-$ state is included). The results 
obtained are in a very good agreement with \cite{ang99}. The increase of the 
rate due to addition of the $5/2^-$ state is $11\%$ at most in the temperature 
range up to 10 GK. This small change is connected with comparatively small gamma 
width of this state:
the gamma widths of the other states in the capture cross section
parametrization used in \cite{ang99} are around $0.45-0.9$ eV. So, the 
uncertainty of the reaction rate due to uncertainties of the experimental data 
found in \cite{ang99} is significantly larger than correction connected with 
$5/2^-$ state (see Table \ref{tab:9be-rate}).

The mentioned experimental uncertainty could be even larger than it was
inferred in Ref.\ \cite{ang99}. The analysis, provided in Ref.\ \cite{efr98} in
the framework of the semimicroscopic model, demonstrated that the older
photodisintegration data for $^9$Be \cite{ham53,gib59,joh62} could be more
preferable than the more up-to-date results \cite{fui82} (on which, e.g.\ the
parametrization of the cross section used in Ref.\ \cite{ang99} is based). The
reaction rate found in paper \cite{efr98} (as well as in the early work
\cite{fow75}) is around $35\%$ larger than the rate in Ref.\ \cite{ang99}.

Paper \cite{buc01} is generally dedicated to the R-matrix analysis of the
$\beta$-delayed particle decay of $^9$C via the excited states in $^9$B. The
authors utilize the R-matrix parameters obtained in the decay studies of $^9$B
for the caption calculations in $^9$Be. The reaction rate calculated in this
work is consistent with the other results at low temperatures, but is
qualitatively different at $T>3$ GK (see Fig.\ \ref{fig:9be-rate}, dotted
curve). The rise of the reaction rate at higher temperatures is connected,
according to \cite{buc01}, with contribution of sequential capture of
$\alpha$-particle on the broad ground state of $^5$He. Such capture path has
never been considered elsewhere. It should be noted that in the framework of
sequential formalism this is a valid question: how narrow should be the
intermediate state, to be considered within this formalism. Really, the
sequential formalism is evidently correct in the limit of infinitely narrow
intermediate state. However, in the other limit (an infinitely broad state), we
have just nonresonant continuum and the sequential formalism should fail at some
point. This issue is qualitatively discussed in Section \ref{sec:formal}. Our
work resolve this question in a very natural way: we state that contributions of
different sequential and three-body channels should add up in a way, which makes
their relative contributions unimportant. So, inclusion of capture via $^5$He
into formalism should not lead to any significant changes (compared to
conventional sequential capture via $^8$Be g.s.), until there exist
states with dominating three-body decay branch (which is not accounted in
sequential formalism) and large gamma widths. No such
states are known in the energy range of interest. The reaction rate from Ref.\
\cite{buc01} can be reproduced within our formalism only if
we assume the gamma width for the 3 MeV state in $^9$Be to be about 15 eV and 
also assume one more state at about 5 MeV with gamma width above 1 keV. Such 
assumptions are quite unrealistic.

Unfortunately, there is an evidence for problems in Ref.\ \cite{buc01}, which
probably have leaded to the discussed strange result. In Eq.\ (32) of this work
the penetrability is present in the first power, while it should be in the
second (as we speak about elastic cross section). Possibly this is the reason of
the qualitatively incorrect behaviour of the intermediate population values (see
Fig.\ 8 in Ref.\ \cite{buc01}). For
example, the population $\langle \sigma (E) v/\Gamma(E)\rangle$ for $^5$He g.s.\
should decrease as $T$ at low temperature. In Fig.\ 8 of Ref.\ \cite{buc01} this
value has a rapid rise at low temperature. Using Eq.\ (32) from Ref.\
\cite{buc01} ``as is'' one gets behaviour $T^{-1/2}$ at low $T$ in agreement
with this figure.

\begin{table}
\caption{Reaction rates multiplied by $N^2_A$ (in cm$^6$/s) for
$^4$He($n\alpha$,$\gamma$)$^{9}$Be reaction.}
\vspace{2mm}
\begin{ruledtabular}
\begin{tabular}[c]{cccc}
$T$ (GK) & Ref.\ \cite{ang99}  & Ref.\ \cite{ang99} upper & This work \\
\hline
 2   & $1.80\times 10^{-7}$ & $2.20\times 10^{-7}$ & $1.83\times 10^{-7}$ \\
 4   & $5.48\times 10^{-8}$ & $6.99\times 10^{-8}$ & $5.94\times 10^{-8}$ \\
 6   & $2.88\times 10^{-8}$ & $3.83\times 10^{-8}$ & $3.20\times 10^{-8}$ \\
 8   & $1.81\times 10^{-8}$ & $2.47\times 10^{-8}$ & $2.02\times 10^{-8}$ \\
 10  & $1.23\times 10^{-8}$ & $1.70\times 10^{-8}$ & $1.36\times 10^{-8}$ \\
\end{tabular}
\end{ruledtabular}
\label{tab:9be-rate}
\end{table}

So, the difference of our approach found in the case of
$^4$He($n\alpha$,$\gamma$)$^{9}$Be reaction is not significant. It is much
smaller than the other uncertainties (see Refs.\ \cite{efr98} and \cite{ang99}).
However, our formalism excludes such a possibility as an importance of broad
intermediate $^5$He state \cite{buc01}.


\section{Conclusion}


We use the formalism based on the S-matrix for $3 \rightarrow 3$ scattering to
derive the reaction rate for the three-body resonant radiative capture. This
derivation makes especially evident that (i) all the three-body states
should be included in the treatment (even if there is no opportunity of a
sequential capture to the state), (ii) the detailed knowledge of the
intermediate
states is unnecessary to calculate the resonant rates and (iii) only the
knowledge of particle and gamma widths for the three-body states is needed to
calculate the resonant rates (not the relative contribution of direct and
sequential mechanisms).

This formalism, together with the modern results on $2p$ and $\gamma$ widths of
$^{17}$Ne states, allows us to update significantly the capture rate for the
$^{15}$O($2p$,$\gamma$)$^{17}$Ne reaction. The updated rate is up to 4--9
orders of the magnitude larger (in the temperature range of astrophysical
interest). The experimental derivation of the $2p$ width of the first excited
state in $^{17}$Ne is found to be very important for refining this rate. The
$^{38}$Ca($2p$,$\gamma$)$^{40}$Ti reaction rate has also got a considerable
increase. Thus the conclusions about importance of the $2p$ capture reactions
could possibly be more optimistic than in Ref.\ \cite{gor95}. We also discuss
the impact of our approach on the $^{18}$Ne($2p$,$\gamma$)$^{20}$Mg, and
$^4$He($n\alpha$,$\gamma$)$^{9}$Be reaction rates. Our studies emphasize the
importance of better gamma width information for $2p$ capture rates
(experimental or theoretical, if the first is not available).

The studies of this work are restricted to resonant reactions (and
correspondingly to relatively high temperatures). We are planning to perform
accurate three-body studies of the nonresonant contributions in the forthcoming
paper.


\begin{acknowledgments}

We are grateful to Prof.\ B.\ V.\ Danilin and Prof.\ N.\ B.\ Shul'gina for 
interesting discussions. LVG was partly supported by Russian RFBR Grant 
02-02-16174 and Ministry of Industry and
Science grant NS-1885.2003.2. We thank E.\ Smirnova for careful reading of the
manuscript.

\end{acknowledgments}



\end{document}